\documentclass[prd,aps,preprint,amsmath,nofootinbib,amssymb,eqsecnum,showkeys]{revtex4}
\pdfoutput=1
\usepackage{verbatim,graphics,graphicx,color,slashed,textcomp}
\usepackage{ulem,soul,xcolor} 
\usepackage{hyperref}

\graphicspath{{figures/}}

\newcommand{\MeV}{\mathrm{MeV}}
\newcommand{\GeV}{\mathrm{GeV}}
\newcommand{\TeV}{\mathrm{TeV}}

\begin{document}
\setstcolor{red}

\title{ 
AMS02 positron excess from decaying fermion DM \\ with local dark gauge symmetry}

\author{P. Ko, Yong Tang}

\affiliation{School of Physics, KIAS \\ Seoul 130-722, Korea}
\date{\today}

\begin{abstract}
Positron excess observed by PAMELA, Fermi and AMS02 may be due to 
dark matter (DM) pair annihilation or decay dominantly into muons. 
In this paper, we consider a scenario with thermal fermionic 
DM ($\chi$) with mass $\sim O(1-2)$ TeV decaying into a dark Higgs 
($\phi$) and an active neutrino ($\nu_a$) instead of the SM Higgs boson 
and $\nu_a$. We first present a renormalizable model for this 
scenario with local dark $U(1)_X$ gauge symmetry, in which
the DM $\chi$ can be thermalized by Higgs portal and the gauge 
kinetic mixing. Assuming the dark Higgs ($\phi$) mass is in the 
range $2 m_\mu < m_\phi < 2 m_{\pi^0}$, 
the positron excess can be fit if a proper background model is used, without
conflict with constraints from antiproton and gamma-ray fluxes 
or direct detection experiments.  Also, having such a light dark Higgs, the self-interaction of DM can be enhanced to some extent, 
and three puzzles in the CDM paradigm can be somewhat relaxed.
\end{abstract}

\maketitle
\flushbottom
\section{Introduction}
\label{sec:intro}
Positron excess in the energy range $E>10\GeV$ has been observed by PAMELA, Fermi and AMS02~\cite{pamela, fermilat, ams02, ams141, ams142}. Assuming its DM-origin 
\footnote{Note that astrophysical processes for this excess are also discussed in Refs. 
~\cite{Gaggero:2013rya, Blum:2013zsa, DiMauro:2014iia, Venter:2014ata, Boudaud:2014dta}.}, 
this excess can be explained by annihilating DM with thermally-averaged cross section 
$\langle \sigma v \rangle \sim 10^{-23}$cm$^3$/s, 
or decaying DM with lifetime $\tau = \Gamma^{-1} \sim 10^{26}$s. It is also well known that for annihilating DM a large boost factor $\sim 10^3$~\cite{Cirelli:2008pk, Baek:2008nz, Bi:2009uj, Chen:2009gz, Pearce:2013ola, DeSimone:2013fia, Yuan:2013eja, Cholis:2013psa, Jin:2013nta, Yuan:2013eba, Yin:2013vaa, Liu:2013vha, Dev:2013hka, Masina:2013yea, Bergstrom:2013jra, Ibarra:2013zia, Lin:2014vja, Jin:2014ica} is needed to fit the positron spectra. 
However, such a large boost factor is strongly constrained by the CMB data~\cite{Padmanabhan:2005es, Galli:2009zc, Slatyer:2009yq, Zavala:2009mi, Madhavacheril:2013cna} and Fermi/LAT gamma ray measurements~\cite{Cirelli:2009dv, Chen:2009uq, Calore:2013yia, Ackermann:2012qk, Ackermann:2012rg, Cirelli:2012ut, Gomez-Vargas:2013bea}. On the other hand, 
$\mathcal{O}(\TeV)$ DM decaying into leptons \cite{Feng:2013zca, Ibe:2013nka, Kajiyama:2013dba, Feng:2013vva, Dienes:2013lxa, Geng:2013nda, Baek:2014goa, Belotsky:2014haa, Ibe:2014qya, Cao:2014cda} can give a consistent explanation without conflict with such stringent constraints, especially for DM decay 
into the $\mu^+ \mu^-$ channel.  Then the remaining question would be to construct particle
physics models for such a scenario.

It is well known that the following operator can fit the positron excess well 
with $\Gamma \sim 10^{-51}$GeV~\cite{Hamaguchi:2008ta}:  
\[
\delta\mathcal{L} = \lambda_{\rm eff} \bar{\chi} \phi \nu,
\]
if $\lambda_{\rm eff} \sim 10^{-26}$, where $\chi$ is the decaying fermion DM, 
$\phi$ is some scalar field and $\nu$ is the SM active neutrino. 

The simplest guess would be to assume $\chi$ is the SM singlet fermion and $\phi$ is the SM Higgs doublet and $\nu$ is from the left-handed lepton doublet~\cite{Hamaguchi:2008ta}. In this case, the operator $\lambda_{\rm eff} \bar{\chi} h \nu$ would also induce $\chi\rightarrow Z \nu$ 
and $\chi\rightarrow W^{\pm}e^\mp$ that would give potentially dangerous antiproton 
or $\gamma$-ray flux. 

In this paper, we focus on the light $\phi$ case ($2m_{e^{\pm}}<m_\phi<2m_{\pi^0}$), 
which has been also motivated to give large self-interaction for DM-DM elastic scattering 
(see Ref.~\cite{Tulin:2013teo} for example). The key observation in this paper is that this
light scalar $\phi$ can be identified as a dark Higgs field which is generically present
in the DM models with local dark gauge symmetries.   
In the following, we shall construct such dark matter models that are renormalizable and 
whose dynamics is completely fixed by local dark gauge symmetry.   In those models, 
one can induce the above dim-4 operator where the dimensionless effective coupling  
$\lambda_{\textrm{eff}}$ is suppressed by heavy mass scales in the intermediate states.  
In our model, $\phi$ will be a new light scalar field (dark Higgs) that would 
eventually decay into light lepton pairs through Higgs portal interaction, and 
$\nu$ is the SM neutrino field.  There is also $Z' \nu$ final state due to the mixing between $\nu$ and $\chi$ in a definite ratio, and we account for it for completeness.

This paper is organized as follows. In Sec.~\ref{sec:model1}, we propose a renormalizable model for decaying fermionic DM $\psi$ based on local $U(1)_X$ gauge symmetry, and show the effective operators generated after dark gauge symmetry breaking. 
In Sec.~\ref{sec:decay}, we discuss the main decay modes of the DM $\psi$, and  several variations of the model are discussed in Sec.~\ref{sec:variant}. 
In Sec.~\ref{sec:numeric}, we compare the theoretical calculations for $e^{\pm}$ spectra with the experimental data from PAMELA, FERMI and AMS02. 
Finally, we make a conclusion in Sec.~\ref{sec:conclusion}.

\section{Model}\label{sec:model1}

We consider local dark gauge symmetry $U(1)_X$ with dark Higgs $\Phi$ and
two different Dirac fermions in the dark sector, $\chi$ and $\psi$.  
Let us assign $U(1)_X$ charges to the dark fields as follows:
\[
(Q_\chi, Q_\psi, Q_\Phi) = (2,1,1) .
\]
Then we can write down all the possible renormalizable interactions
including singlet right-handed neutrinos $N$ for the neutrino masses and mixings: 
\begin{align}
\mathcal{L}=& \mathcal{L}_{\mathrm{SM}} + \frac{1}{2}\bar{N_I}i\slashed{\partial}N_I - 
\left(\frac{1}{2}m_{NI} \bar{N}^c_I N_I + y_{\alpha I} \bar{L}H N_I  + \textrm{h.c.}\right)\nonumber \\
&-\frac{1}{4}X_{\mu\nu}X^{\mu\nu} -\frac{1}{2}\sin{\epsilon}X_{\mu\nu}F^{\mu\nu}_{Y}  + \left(D_\mu \Phi \right)^{\dagger}D^\mu \Phi -V(\phi,H)\nonumber\\
& + \bar{\chi}\left(i\slashed{D} - m_\chi\right)\chi + \bar{\psi}\left(i\slashed{D} - m_{\psi}\right)\psi  - \left(f \bar{\chi}\Phi \psi + g_I \bar{\psi}\Phi N_I  +\textrm{h.c.}\right), 
\end{align}
where $L_\alpha = (\nu_\alpha \; l_\alpha)^T$ is the left-handed SM $SU(2)$ lepton 
doublet with $\alpha=e,\mu,\tau$. $H$ is the SM Higgs doublet, $X_{\mu\nu} \equiv 
X_{\mu\nu}=\partial_\mu X_\nu - \partial_\nu X_\mu$ is the field strength tensor of $U(1)_X$ dark gauge field $X_\mu$, $F^{\mu\nu}_Y$ is for SM hypercharge $U(1)_Y$, and $\epsilon$ is the kinetic mixing parameter. Covariant derivative is defined as
\[
D_\mu C = \left(\partial_\mu - i g_X Q_C X_\mu \right) C, 
\ \ \  ( C=\chi,\psi,\Phi ).
\]
and the scalar potential is given by 
\begin{equation}
V=\lambda_H \left(H^\dagger H -\frac{v^2_H}{2}\right)^2+\lambda_{\phi H}\left(H^\dagger H -\frac{v^2_H}{2}\right) \left(\Phi^\dagger \Phi -\frac{v^2_\phi}{2}\right) + \lambda_\phi \left(\Phi^\dagger \Phi -\frac{v^2_\phi}{2}\right)^2.
\end{equation}
To explain the neutrino oscillation, at least two RH neutrinos $N$'s are introduced 
in order to generate two non-zero neutrino masses. However, for our study of positron 
excess, we can focus only on the case with one $N$ without loss of generality. 
Therefore, we shall omit the lower indices for $N_I$, $L_\alpha$, $m_N$, 
$y_{\alpha I}$ and $g_I$ from now on. 

\begin{figure}[t]
\centering
\includegraphics[width=0.56\textwidth,height=0.2\textwidth]{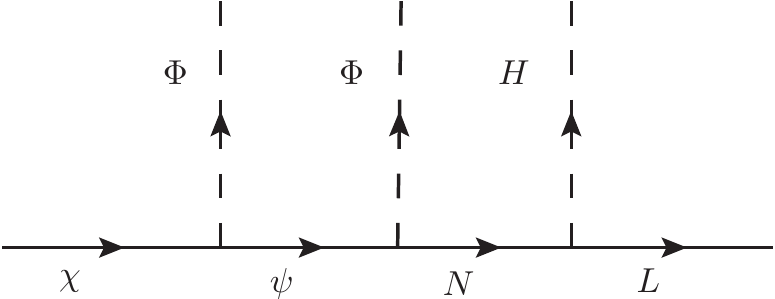}
\caption{Feynman diagram that generates the effector operator $\bar{\chi}\Phi \Phi \tilde{H} L$.}
\label{fig:effective}
\end{figure}

We are interested in the case where  $m_\chi\sim \TeV$ and 
$v_\phi\sim \mathcal{O}(100)\MeV$ while $m_N$ and $m_\psi$ are much heavier. 
Integrate both $\psi$ and $N$,  we get an interesting dim-6 operator: 
\begin{equation}\label{eq:effective}
\frac{yfg}{m_\psi m_N}\bar{\chi}\Phi \Phi \tilde{H} L \ .
\end{equation}
Diagrammatically, it can be represented as the Feynman diagram in 
Fig.~\ref{fig:effective}.

The local gauge symmetry of this model is broken by the following vacuum 
configurations: 
\begin{eqnarray}\label{eq:vacuum}
\langle H\rangle=\frac{1}{\sqrt{2}}\left(\begin{array}{c}
0\\
v_{H}
\end{array}\right),\;\langle\Phi\rangle=\frac{v_{\phi}}{\sqrt{2}} \ ,
\end{eqnarray}
where $v_H\simeq 246$GeV and $v_{\phi}\sim \mathcal{O}(100)$MeV for our interest. $v_{\phi}\sim \mathcal{O}(100)$MeV is motivated for having light mediators in the dark sector such that they can only decay into $e^\pm/\mu^\pm$ and provide a large DM self-scattering cross section~\cite{Ko:2014bka}. The model Lagrangian (2.1) is basically the same as the one discussed in Ref.~\cite{Ko:2014bka} by the present authors, except that the hidden sector fermions carry definite $U(1)_X$ charges and one of them $\psi$ is very heavy $\sim 10^{14}$GeV in this work.

In the unitarity gauge, we can replace the scalar fields with
\begin{equation}
 H\rightarrow \frac{1}{\sqrt{2}}
 \left(
	 \begin{array}{c}
		 0 \\
		 v_{H}+h(x)
	 \end{array}
 \right)
  ~~ {\rm and}~~
 \phi_X\rightarrow\dfrac{v_\phi+\phi(x)}{\sqrt{2}},
\end{equation}
where $h$ and $\phi$ are two real scalar fields which mix with each other because 
of the Higgs-portal interaction, $\lambda_{\phi H}H^\dagger H\Phi^\dagger \Phi$. 
Through this mixing, dark Higgs $\phi$ can decay into SM particles. Another mixing is concerned with three neutral gauge bosons, photon $A_\mu$, $Z_\mu$ and $X_\mu$. Such a mixture enable an extra mass eigenstate $Z_{\mu}'$(mostly $X_\mu$) to decay SM fermion pairs. Note that the dark Higgs boson decays dominantly into heavier particles, 
thus being naturally flavor dependent, unlike the dark photon $Z^{'}$.  
DM $\chi$'s scattering off nucleus then is possible by exchanging a $\phi$ or $Z'$, 
whose cross section depends on $\lambda_{\phi H},\epsilon,v_\phi, m_\phi, m_{Z'}$. 
It is easy to choose these parameters and evade the stringent constrains from DM direct detection, see Ref.~\cite{Baek:2014goa} for example. 

Typically, for $m_{Z'}\sim \mathcal{O}(100)\MeV$, the kinetic mixing parameter $\epsilon$ should be around $[10^{-10},10^{-7}]$, where the upper and the lower bounds come from low energy beam dump experiments~\cite{Bjorken:2009mm} and from BBN and supernovae constraints~\cite{Dent:2012mx}, respectively. On the other hand, the Higgs portal coupling $\lambda_{\phi H}$ can be much larger than $\epsilon$.  $\lambda_{\phi H}$ in the range $10^{-7}\lesssim \lambda_{\phi H}\lesssim 10^{-3}$ would be small enough to give $Br(h\rightarrow \phi \phi)\lesssim 2\%$, but sufficiently large to thermalize the dark 
sector around $\TeV$ in the early Universe.

After spontaneous gauge symmetry breaking, we have several dimensional effective operators as follows: 
\begin{align}
\textrm{dim-}3:\;& \frac{v^2_\phi v_H}{m_\psi m_N}\bar{\chi}\nu \ ,\label{eq:dim3}\\
\textrm{dim-}4:\;&\frac{v^2_\phi}{m_\psi m_N}\bar{\chi}h\nu \ , \;\;~
\frac{2v_\phi v_H}{m_\psi m_N}\bar{\chi}\phi \nu,\label{eq:dim4}\\ 
\textrm{dim-}5:\;&\frac{v_H}{m_\psi m_N}\bar{\chi}\phi\phi\nu \ ,\;
\frac{2v_\phi}{m_\psi m_N}\bar{\chi}\phi h\nu,\label{eq:dim5}\\
\textrm{dim-}6:\;&\frac{1}{m_\psi m_N}\bar{\chi}\phi\phi h\nu \ . \label{eq:dim6} 
\end{align}
omitting the common factor $\dfrac{yfg}{4\sqrt{2}}$. 

Discussion of $\bar{\chi}h\nu$ operator has been presented in Ref.~\cite{Hamaguchi:2008ta} in great detail.   
As we shall see in the next section, it is the operator $\bar{\chi}\phi\nu$ rather than $\bar{\chi}h\nu$ that gives the dominant contribution to the positron flux in our model, if we assume $m_\phi \ll m_H$. Then our model does not suffer much from the constraints from antiproton and gamma-ray fluxes on the dark matter decays. 

\section{Decay Modes}\label{sec:decay}

\begin{figure}[t]
\centering
\includegraphics[width=0.34\textwidth]{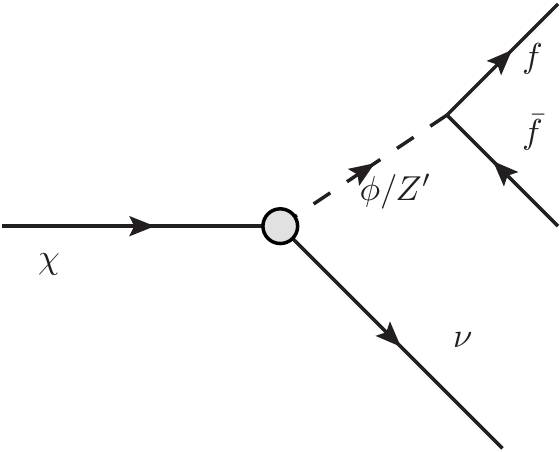}
\caption{Dominant decaying process.}
\label{fig:decay}
\end{figure}

The dim-3 operator, Eq.~(\ref{eq:dim3}), is a mass term and would induce a tiny 
mixing between $\chi$ and $\nu$ with  the mixing angle, 
\begin{equation}
\beta\simeq \frac{yfg}{4\sqrt{2}}\frac{v^2_\phi v_H}{m_\psi m_N m_\chi} \ . 
\end{equation}
Then the gauge interactions for $\chi$ and $\nu$ will generate the decay channels,
\begin{equation}
\chi\rightarrow Z'\nu, Z\nu, l^{\pm}W^{\mp},
\end{equation}
with their branching ratios being proportional to  $\sim v^2_H:v^2_\phi:v^2_\phi$.

Two dim-4 operators, Eq.~(\ref{eq:dim4}), lead to $\chi$ decays,
\begin{equation}
\chi\rightarrow h\nu,\phi \nu,
\end{equation}
with their branching ratios being proportional to $\sim v^2_\phi : 4v^2_H$. 
Since $m_h \gg m_\phi$ would generically imply $v_H\gg v_\phi$, we would expect 
$\Gamma_{\chi \rightarrow \phi \nu}\gg \Gamma_{\chi \rightarrow h \nu}$.
It is also straightforward to get the following relation for the branching ratios, 
\begin{equation}
Br(\chi \rightarrow \phi \nu):Br(\chi \rightarrow Z' \nu)= 2^2:1 \ .
\end{equation}
The factor $2^2$ results from $2$ in the numerator of the second operator in Eq.~(\ref{eq:dim4}), which stems from two $\Phi$'s in the dim-6 operator in Eq.~(\ref{eq:effective}). On-shell $\phi/Z'$ then decay into light SM fermion 
pair, as shown in Fig.~\ref{fig:decay}.

In this model, we can estimate 
\begin{equation}
\lambda_{\rm eff} \sim \frac{yfg}{4\sqrt{2}}\frac{v_\phi}{m_\psi} \frac{v_H}{m_N} \sim 10^{-26} \ ,
\end{equation}
which can be easily achieved if we choose model parameters as
\begin{equation}
v_\phi \sim O(100) {\rm MeV} \ ,\; m_N \sim m_\psi \sim 10^{14} {\rm GeV} 
\ ,\; yfg\sim 1 \ .
\end{equation}

Finally, a dim-5 operators, Eq.~(\ref{eq:dim5}), would induce three-body decay 
channels, $\chi\rightarrow \phi \phi \nu$ and $\chi\rightarrow \phi h \nu$. 
These decays are, however, less dominant than two-body decays considered earlier   
if $m_\chi \lesssim 3$TeV, because of  
\begin{equation}
\dfrac{\Gamma_{3\textrm{-body}}}{\Gamma_{2\textrm{-body}}}\propto \frac{1}{\left(4\pi\right)^2}\dfrac{m^2_\chi}{v^2_H}.
\end{equation}
The $1/(4\pi)^2$ suppression factor comes from the phase space integration. Four-body decay $\chi\rightarrow \phi\phi h \nu$ is then even suppressed for $m_\chi \lesssim 3$TeV. 
Also, even for $m_\chi > 3$TeV,  the positron spectrum resulting from those multiple final 
states are softer that the two-body decay case and we shall not consider their 
contributions in this paper.

\section{Variant models}\label{sec:variant}

\begin{figure}[t]
\centering
\includegraphics[width=0.45\textwidth,height=0.2\textwidth]{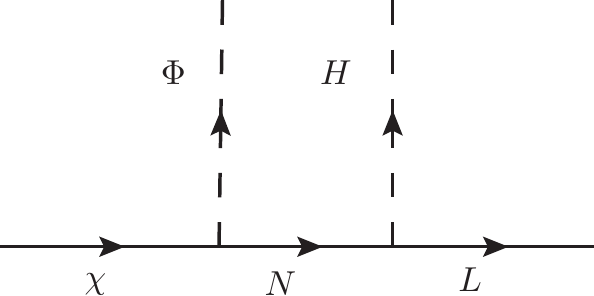}
\caption{Feynman diagram that generates the effector operator $\bar{\chi}\Phi \tilde{H} L$.}
\label{fig:effective2}
\end{figure}

\begin{figure}[t]
\centering
\includegraphics[width=0.55\textwidth,height=0.2\textwidth]{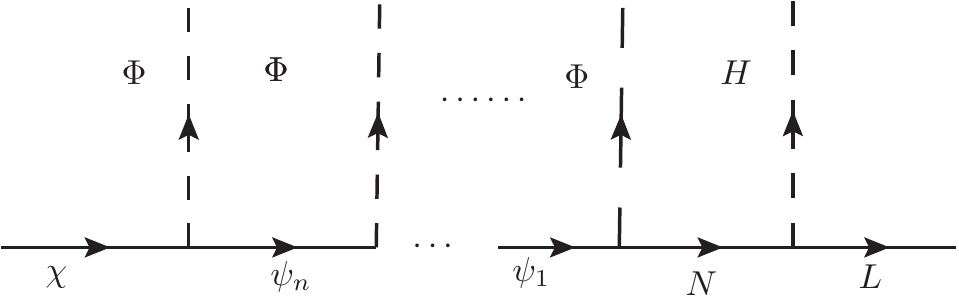}
\caption{Feynman diagram that generates the effector operator $ \bar{\chi} \Phi^{n+1}\tilde{H}L$.}
\label{fig:effective3}
\end{figure}

One can consider some variations of the model discussed in the previous section,
by modifying the $U(1)_X$ charge assignments to the dark fields, thereby changing
the relative branching ratios of the DM decays into $\phi + \nu$ and $H + \nu$.

Let us first consider the following assignments: 
\begin{equation}
\left(Q_\chi,Q_\Phi\right)=(1,1). 
\end{equation}
Then we can have Yukawa interaction term $f\bar{\chi}\Phi N$, 
and we do not need $\psi$ to induce $\chi$ to decay. 
However, in this case, we need tiny Yukawa couplings. 
Integrating out the heavy RH neutrino $N$ in Fig.~\ref{fig:effective2},  
the following dim-5 operators would be generated:
\begin{equation}
\frac{yf}{m_N}\bar{\chi}\Phi \tilde{H}L. 
\end{equation}
Then the effective $\lambda_{\textrm{eff}}$ is
\begin{equation}
\lambda_{\textrm{eff}}\sim \frac{yf}{2}\frac{v_H}{m_N}\sim 10^{-26}.
\end{equation}
In this model we have a different branch ratio, 
\begin{equation}
Br(\chi \rightarrow \phi \nu):Br(\chi \rightarrow Z' \nu)= 1:1.
\end{equation}

If the dark symmetry were global rather than local, then we would not have the dark gauge boson $Z'$, and correspondingly $Br(\chi \rightarrow \phi \nu)\simeq 1$. However, in this case, in the early Universe $\chi$ would not be thermalized at $\TeV$ in the minimal setup and we may also need to add new fields to deal with the Goldstone mode, which is beyond our discussion in this paper. 

From the previous discussion, it is easy to see that we can generalize the above mechanism with $n$ low-scale $\psi$'s by assigning the $U(1)_X$ charges as
\begin{equation}
(Q_\chi, Q_{\psi_n},...,Q_{\psi_1},\Phi)=(n+1,n,...,1,1).
\end{equation}
Then the following effective operator will be generated,
\begin{equation}
\frac{yg }{(n+1)! m_N}\frac{f_n\cdots f_1}{m_{\psi_n}\cdots m_{\psi_1}} \bar{\chi} \Phi^{n+1}\tilde{H}L.
\end{equation}
Feynman diagram is shown in Fig.~\ref{fig:effective3}. In this case branching ratio for our interest would be 
\begin{equation}
Br(\chi \rightarrow \phi \nu):Br(\chi \rightarrow Z' \nu)= n^2:1.
\end{equation}

\section{Positron Fraction and Flux}\label{sec:numeric}

In this section we calculate the $e^{\pm}$ flux ($\Phi_{e^{\pm}}$) on earth.  
It is the sum of two contributions from DM decay \footnote{In principle both 
$\chi \overline{\chi}$ pair annihilation and $\chi$ decay can give rise to $e^\pm$.  
However in our interested parameter ranges, we have checked that $\chi$ decay is 
the dominant one even taking the enhancement factor into account for $\chi$ pair 
annihilation. Therefore we shall only focus on the signature from $\chi$ decays.} 
and astrophysical background, $\Phi_{\pm}=\Phi_{e^\pm}^{\mathrm{DM}}+\Phi_{e^\pm}^{\mathrm{bkg}}$, and will compare
with the experimental observation. 
We use \texttt{PPPC4DMID}~\cite{pppc4dmid} to compute $\Phi_{e^\pm}^{\mathrm{DM}}$, 
and adopt the Einasto density profile for DM halo profile~\cite{Einasto}:  
\begin{equation}
\rho_{\mathrm{DM}}=\rho_s\textrm{exp}\left[-\frac{2}{\alpha}\left(\left[\frac{r}{r_s}\right]^\alpha-1\right)\right],
\end{equation}
where $\alpha=0.17$, $r_s=28.44$kpc and $\rho_s=0.033$GeV/cm$^3$. And the $e^{\pm}$ background $\Phi_{e^\pm}^{\mathrm{bkg}}$  is taken from Ref.~\cite{Lin:2014vja},  where 
the background electron and positron fluxes $\Phi_{e^\pm}^{\mathrm{bkg}}$ were calculated 
by assuming the injection spectra of all kinds of nuclei with one break, and the injection 
spectrum of primary electron with two breaks, respectively~\footnote{We thank the anonymous referee to point out this to us.}.

As shown in previous section, in our model the dominant decay modes are 
$\chi\rightarrow \phi\nu$ and $\chi\rightarrow Z'\nu$, so that the total decay width can be 
approximated by 
\begin{equation}
\Gamma \simeq \Gamma\left(\chi\rightarrow \phi\nu\right)+\Gamma\left(\chi\rightarrow Z'\nu\right).
\end{equation}
We choose $m_{Z'}$ and $m_\phi$ to lie in the range, $2m_\mu<m_{Z'/\phi}<2m_{\pi^0}$, 
such that the available final states in the $Z'/\phi$  decays are $\nu\bar{\nu}, e^\pm$ and 
$\mu^\pm$ only.  

While light dark Higgs boson $\phi$ mainly decays into $\mu^-\mu^+$,  light dark photon 
$Z'$ has a quite different decay pattern. In the limit of small  kinetic mixing $\epsilon$, 
the couplings for $Z_{\mu}'\bar{f}_R\gamma^\mu f_R$ and 
$Z_{\mu}'\bar{f}_L\gamma^\mu f_L$ are given by 
\begin{eqnarray}
   &\left[g_1 (s_W-1) Y\right]\sin\epsilon, \\
   &\left[g_1 (s_W-1) Y - g_2 c_W T_3\right]\sin\epsilon,
\end{eqnarray}
respectively.   Here $s_W(c_W)\equiv \sin \theta_W(\cos \theta_W)$, $\theta_W$ is the 
weak mixing angle,  $Y$ is the $U(1)_Y$ hypercharge and $T_3$ is the 3rd component of 
the  $SU(2)_L$ weak isospin generators. 
Then the branching ratios of $Z'\rightarrow \nu \bar{\nu}, e^-e^+$ and $\mu^-\mu^+$ are
\begin{equation}
Br(\nu\bar{\nu}):Br(e^-e^+):Br(\mu^-\mu^+)=\dfrac{3}{4 + \left[1 - \dfrac{g_2 c_W}{2g_1 (s_W-1)}\right]^2}:1:1\simeq 0.7:1:1.
\end{equation}
We include these final states when we calculate the positron flux. 

\begin{figure}[t]
\centering
\includegraphics[width=0.55\textwidth,height=0.50\textwidth]{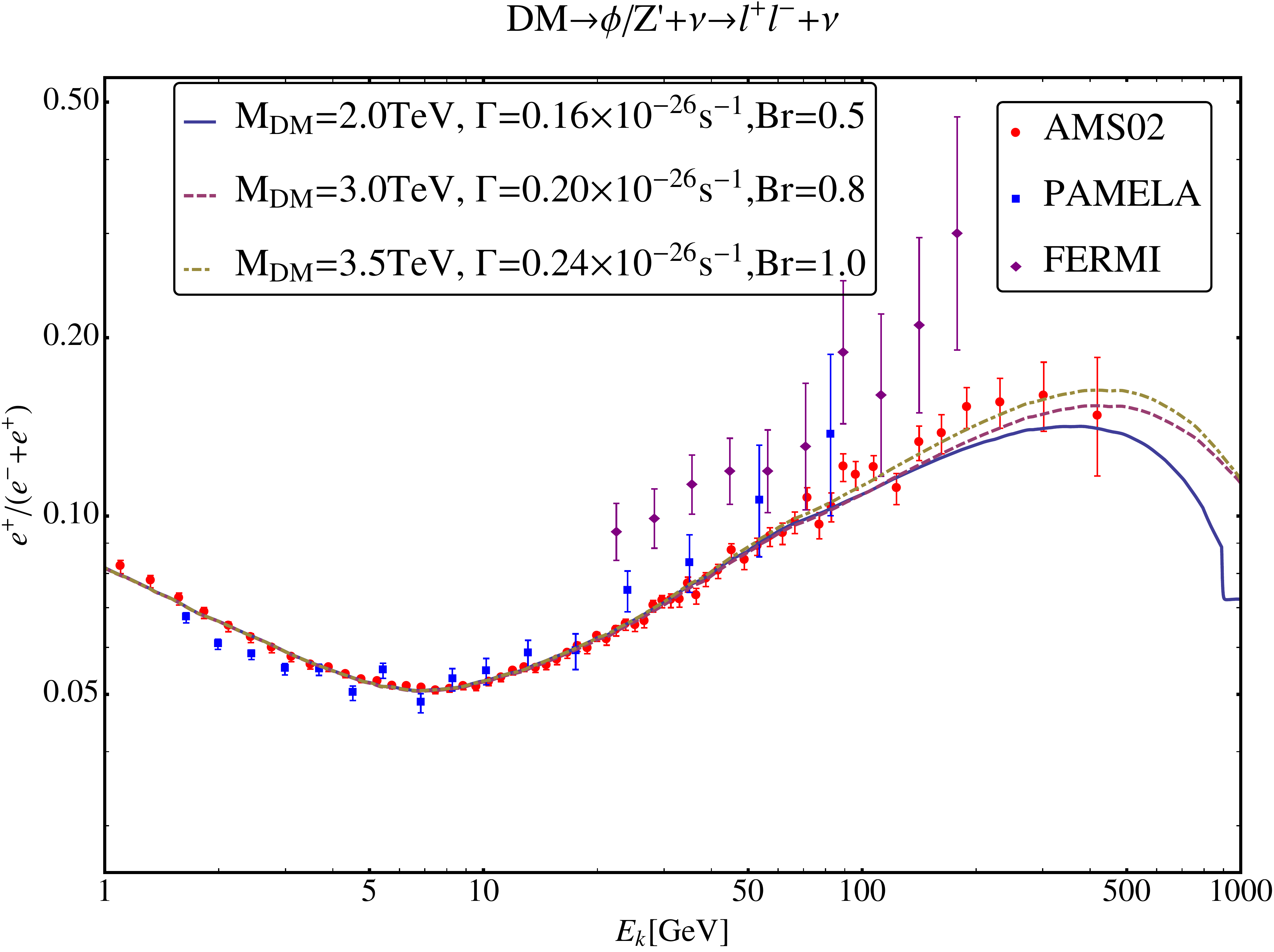}
\caption{Positron fraction in three different sets of parameters. $M_{\textrm{DM}}$ and total 
decay width $\Gamma$ are chosen to visually match the positron fraction data. 
Data are extracted from Ref.~\cite{Maurin:2013lwa}.}
\label{fig:ams}
\end{figure}

Defining the branching ratio as $Br\equiv\Gamma\left(\chi\rightarrow \phi\nu\right)/\Gamma$, 
we illustrate the positron fractions and the fluxes for  following three cases in Fig.~\ref{fig:ams},
\begin{align}
1:\; & M_{\mathrm{DM}}=2.0\TeV,\; \Gamma=0.16\times 10^{-26}s^{-1},\; Br=0.5,\\
2:\; & M_{\mathrm{DM}}=3.0\TeV,\; \Gamma=0.20\times 10^{-26}s^{-1},\; Br=0.8,\\
3:\; & M_{\mathrm{DM}}=3.5\TeV,\; \Gamma=0.24\times 10^{-26}s^{-1},\; Br=1.0.
\end{align} 
Cases 1 and 2 correspond to the effective operators with local gauge symmetry, $\chi \Phi \tilde{H}L$ and $\chi \Phi \Phi\tilde{H}L$, respectively, whereas case 3 corresponds to the 
$\chi \Phi \tilde{H}L$ with global symmetry or $\chi \Phi^n \tilde{H}L$ when $n$ is very large (see Eq. (4.7)). 

\begin{figure}[t]
\includegraphics[width=0.49\textwidth,height=0.50\textwidth]{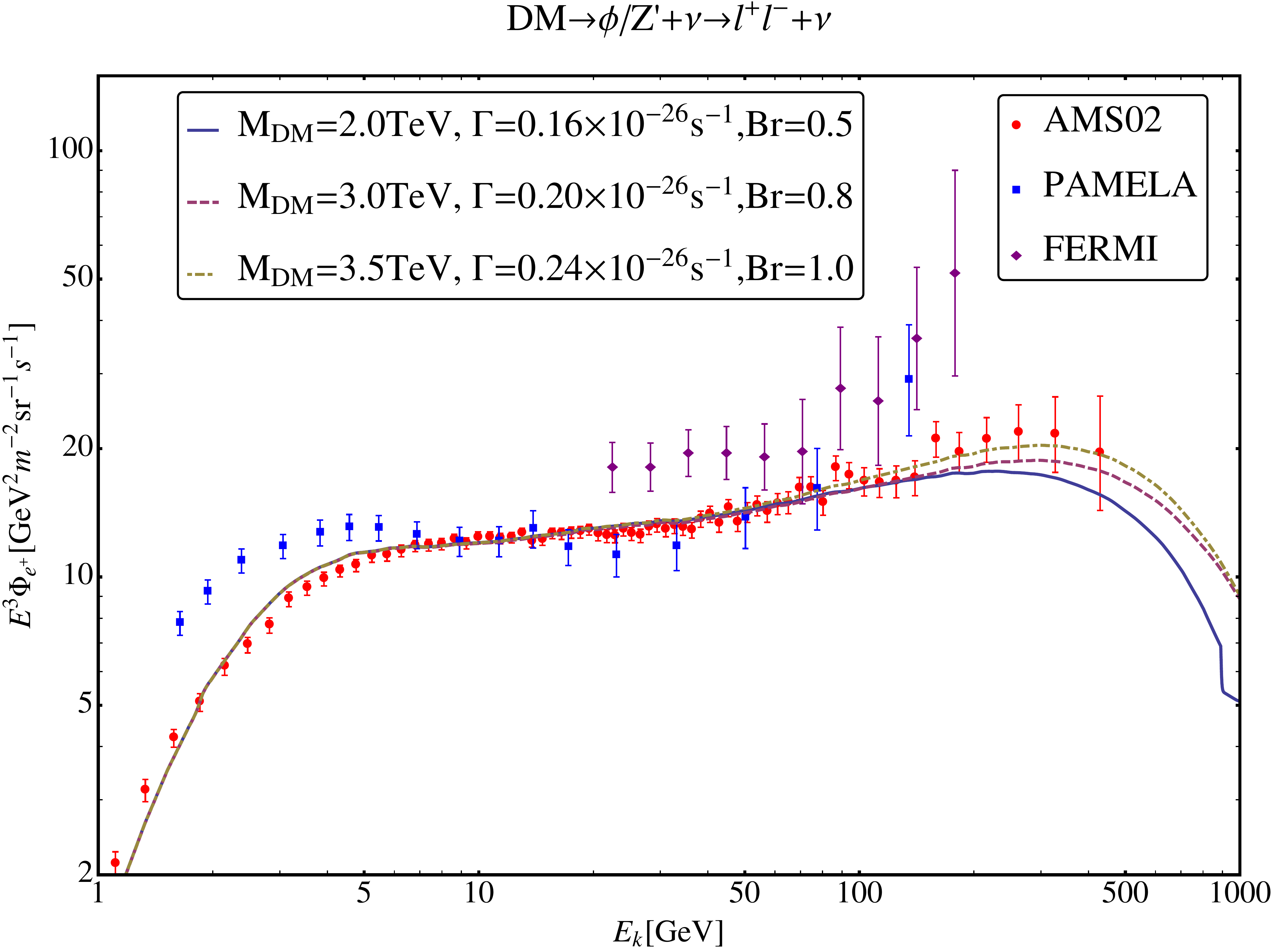}
\includegraphics[width=0.49\textwidth,height=0.50\textwidth]{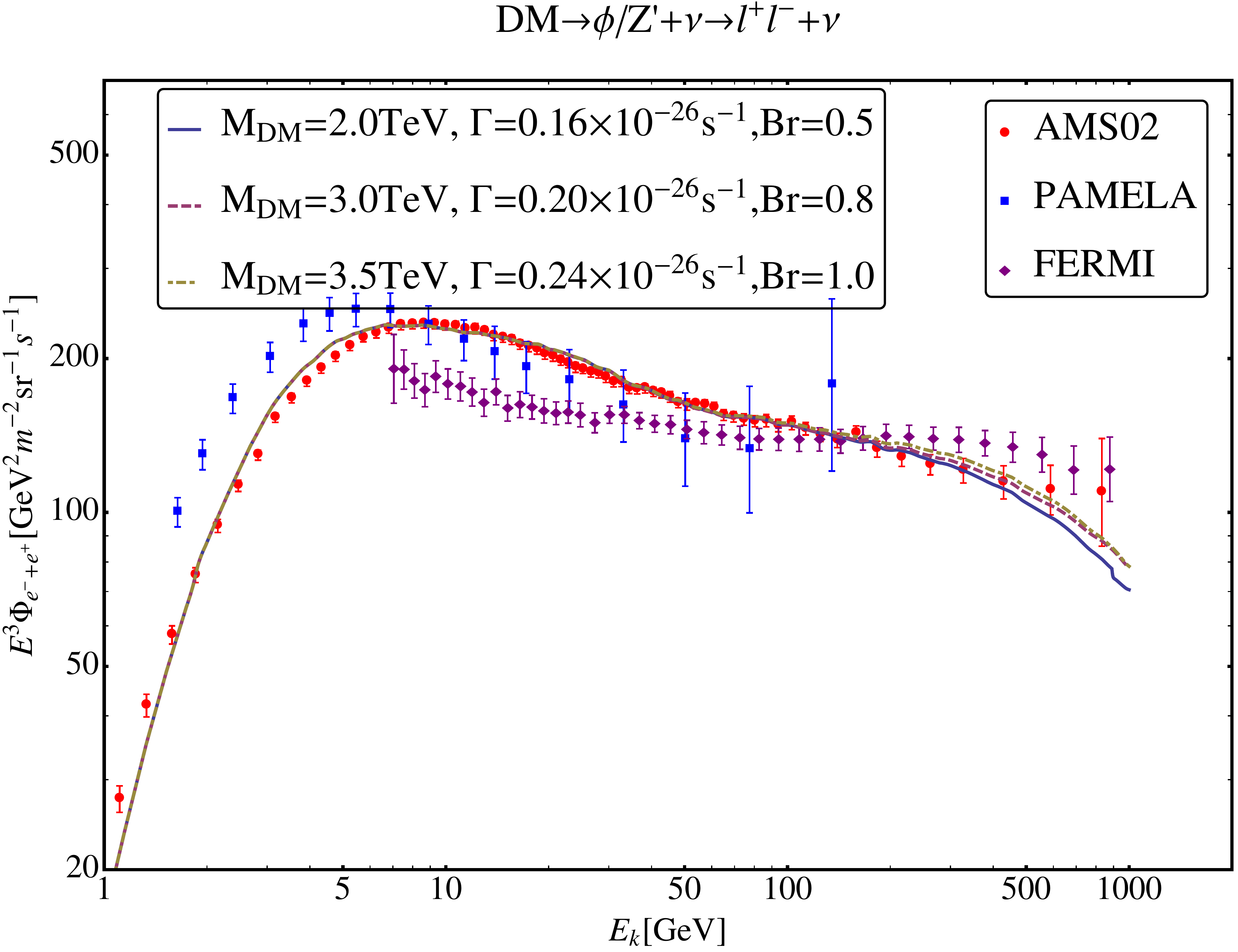}
\caption{Positron flux (left) and electron$+$positron flux (right)~\cite{Aguilar:2014fea,Ackermann:2010ij,Adriani:2013uda} for three different sets of parameters described in the text, Eqs. (5.6)-(5.8).}
\label{fig:flux}
\end{figure}

As shown in Fig.~\ref{fig:ams} for positron fraction, when the branching ratio of $\chi\rightarrow \phi \nu$ increases, we need to increase the DM mass $M_{\textrm{DM}}$ and decay width $\Gamma$ too. 
This feature can be easily understood as follows. Since $Z'\rightarrow e^{+}e^{-}$ gives  harder $e^\pm$ spectra than $\phi\rightarrow \mu^+\mu^-$ does, decreasing the contribution of $Z'\rightarrow e^{+}e^{-}$ would need to be compensated by larger $M_{\textrm{DM}}$ and $\Gamma$. 

For completeness, we also show the positron flux $\Phi_{e^+}$ and the electron$+$positron total flux $\Phi_{e^- + e^+}$ in Fig.~\ref{fig:flux} with the same sets of parameters chosen above.
Note that there is no considerable difference in three cases we considered, except in the high energy regime $\gtrsim 500$GeV. Since $\mu^{+}\mu^{-}$ is the dominant channel 
($\mu^{+}\mu^{-}:e^{+}e^{-}\gtrsim 3.7:1$),  we would expect that all cases can give 
reasonable fits to both $\Phi_{e^+}$ and $\Phi_{e^- + e^+}$. 

Since our discussions are focused in the mass range, $2m_\mu<m_{Z'/\phi}<2m_{\pi^0}$, 
there is no hadronic decay modes for $Z'/\phi$. Then it would not generate additional antiproton flux. The potential constraints come from the $\gamma$-ray flux which are generated by the $e^\pm$ and $\mu^\pm$.
It is expected that constraint  would be more stringent for smaller $Br(\chi\rightarrow \phi \nu)$, since $e^\pm$ gives larger $\gamma$-ray flux than $\mu^\pm$ does. The constraint from the $\gamma$-ray, especially from the galaxy center region in case of DM pair 
annihilation, is also largely dependent on the assumed DM density profile. For example, the 
gamma-ray constraint from the galaxy center will exclude the preferred region if NFW profile is assumed~\cite{Lin:2014vja}. However, the bound could be much weaker if a flatter Einasto-like 
profile is used. And the $\gamma$-ray constraint is even weaker for decaying dark matter 
scenario (see Ref.~\cite{Meade:2009iu} for comparison for example).  Therefore, in our scenario
with decaying DM for AMS02 positron excess, the $\mu^\pm$-channel should be safely allowed.

\section{Conclusion}\label{sec:conclusion}
In this paper, we have proposed a decaying fermionic thermal dark matter model with local 
$U(1)_X$ dark gauge symmetry that can explain the positron excess  together with a proper background model through its decay into 
a light dark Higgs and an active neutrino, rather than into the SM Higgs boson and active 
neutrino.   After integrating the heavy states, $\psi$ and $N$, an effective operator 
\[
\bar{\chi}\Phi\Phi \tilde{H}L
\] 
is generated. Once gauge symmetry is broken spontaneously, we have $\bar{\chi} \phi \nu$ 
which induces the DM $\chi$ to decay into $\phi + \nu$. And $\chi$'s long lifetime $\sim 10^{26}s$ can be easily achieved  when heavy particles have mass around $10^{14}\GeV$. More general mechanism to generate operators $\bar{\chi}\Phi^n \tilde{H}L$ with integer $n$ was also presented. 
We then illustrated with several cases in which the positron fraction and flux spectra can match the experimental data well. Assuming dark Higgs and dark photon are below the dipion threshold $2 m_{\pi^+}$, we could evade the stringent bounds from antiproton and gamma-ray flux measured by Fermi/LAT and other collaborations.  This has more advantage compared with the model where fermionic DM decays into the SM Higgs boson and active neutrinos.

\begin{acknowledgments} 
We are grateful to Seungwon Baek for useful discussions. 
This work is supported in part by National Research Foundation of Korea (NRF) Research 
Grant 2012R1A2A1A01006053, and by the NRF grant funded by the Korea 
government (MSIP) (No. 2009-0083526) through  Korea Neutrino Research Center 
at Seoul National University (P.K.).
\end{acknowledgments}


\end{document}